\documentclass{Interspeech}

\interspeechcameraready 

\title{VS-Singer: Vision-Guided Stereo Singing Voice Synthesis with Consistency Schrödinger Bridge
\thanks{*Corresponding author: Hao Huang}
\thanks{This work was supported by National Natural Science Foundation of China (62466055).}
}

\author[affiliation={1}]{Zijing}{Zhao}
\author[affiliation={1,2}]{Kai}{Wang}
\author[affiliation={1,2,3,*}]{Hao}{Huang}
\author[affiliation={1,2}]{Ying}{Hu}
\author[affiliation={1,4}]{Liang}{He}
\author[affiliation={5}]{Jichen}{Yang}

\affiliation{School of Computer Science and Technology}{ Xinjiang University}{China}
\affiliation{}{Xinjiang Key Laboratory of Multi-lingual Information Technology}{China}
\affiliation{}{Joint International Research Laboratory of Silk Road Multilingual Cognitive Computing}{China}
\affiliation{Department of Electronic Engineering}{Tsinghua University}{China}
\affiliation{School of Cyber Security}{ Guangdong Polytechnic Normal University}{China}
\email{mirror\_zhao@stu.xju.edu.cn,huanghao@xju.edu.cn}

\usepackage{comment}
\usepackage{cite}
\usepackage{hyperref}
\usepackage{amsmath,amssymb,amsfonts}
\usepackage{algorithmic}
\usepackage{graphicx}
\usepackage{textcomp}
\usepackage{xcolor}
\usepackage{multirow}
\usepackage{bm} % 添加这一行以启用 \bm{}
\usepackage[redeflists]{IEEEtrantools}
\def\BibTeX{{\rm B\kern-.05em{\sc i\kern-.025em b}\kern-.08em
    T\kern-.1667em\lower.7ex\hbox{E}\kern-.125emX}}

\begin{document}

\maketitle

\begin{abstract}
    
To explore the potential advantages of utilizing spatial cues from images for generating stereo singing voices with room reverberation, 
we introduce VS-Singer, a vision-guided model designed to produce stereo singing voices with room reverberation from scene images. 
% VS-Singer comprises a modal interaction network (MIN), a decoder based on consistency Schrödinger bridge (CSB), and a spatially-aware feature enhancement module (SFE). Specifically, the MIN integrates spatial features into text encoding to create a linguistic representation enriched with spatial information. Following this, the decoder employs the CSB to construct a manageable diffusion bridge between the representation and binaural mel-spectrograms, facilitating one-step sample generation. 
VS-Singer comprises three modules: firstly, a modal interaction network integrates spatial features into text encoding to create a linguistic representation enriched with spatial information. Secondly, the decoder employs a consistency Schrödinger bridge to facilitate one-step sample generation. 
Moreover, we utilize the SFE module to improve the consistency of audio-visual matching. To our knowledge, this study is the first to combine stereo singing voice synthesis with visual acoustic matching within a unified framework. Experimental results demonstrate that VS-Singer can effectively generate stereo singing voices that align with the scene perspective in a single step.
\end{abstract}

\section{Introduction}
Singing voice synthesis (SVS) involves generating singing voice from musical scores and lyrics, which include features such as notes and pitch \cite{liu2022diffsinger,zhang2022visinger,huang2024audiogpt,gao2024end,lei2023unisyn}. As the continuous development of diffusion models \cite{karras2022elucidating,ho2022video,DBLP:Liu00ZZ22,blattmann2023align,gal2022image,ruiz2023dreambooth,peebles2023scalable}, the naturalness and fluency of synthesized singing speech have now approached those of human performances. Grad-TTS \cite{popov2021grad} introduces stochastic differential equation (SDE) and uses a numerical ordinary differential equation (ODE) solver \cite{song2021scorebased} to solve the reverse SDE, but this reverse process requires thousands of iterations to produce high-quality speech. DiffSinger \cite{liu2022diffsinger} employs a shallow diffusion mechanism, using an auxiliary acoustic model to generate mel-spectrograms, which reducing the sampling process to just one hundred steps. While these models can generate high-quality audio, the excessive number of sampling iterations results in slow inference speed. To alleviate this issue, CoMoSpeech \cite{ye2023comospeech} first pretrains an acoustic model as teacher model, then performs consistency distillation \cite{song2023consistency} using the pretrained model \cite{ren2020fastspeech}, allowing speech samples to be generated in just one sampling step. 
However, this two-stage training process significantly increases training costs, and if the consistency model \cite{song2023consistency} is trained independently without the pre-trained teacher model, its performance declines drastically.

The primary focus of current research in SVS is to improve the naturalness and fluency of synthesized singing voice. However, with the advancement of AR/VR, users have raised their expectations, seeking a more immersive auditory 
% \begin{figure}
% \begin{minipage}[t]{1\linewidth}
%   \centering
%   % \includegraphics[width=0.75\textwidth,height=0.41\textwidth]{wer_result.jpg}
%   \centerline{\includegraphics[width=8cm]{viewpoints_v3.png}}
% %  \vspace{2.0cm}
% \end{minipage}
% %   \centering
% %    \includegraphics[width=0.5\linewidth]{}
%     \caption{Vision-Guided Stereo Singing Voice Synthesis.}
%      % \vspace{-0.1cm}
%     \label{fig:pipelines}
%     % \vspace{-0.6cm}
% \end{figure}
experience.  To achieve this, mainstream research typically converts mono audio into stereo audio through spatial processing. Some researchers \cite{gao20192,lim2024enhancing,liu2025icassp}  proposed matching visual and audio features using attention mechanisms to convert audio into sounds that appear to be recorded in the target environment.
Building on this, NVAS \cite{chen2023novel} designed a network to analyze audio-visual features and convert audio observed from a given perspective into audio for the target perspective. 
 Liu et al. \cite{liu2024visually} introduced the use of masking layers to create left and right eye views that match the observer's perspective, enhancing the audio in the corresponding channels. 
 M2SE-VTTS \cite{liu2025aaai} proposes to simultaneously use the RGB and Depth spaces of spatial images to model local and global spatial knowledge.
However, there has not yet been an effort to research stereo singing voice synthesis.

To bridge this gap and investigate the potential benefits of leveraging spatial cues from images to generate stereo singing voices with room reverberation, this work proposes a novel method named VS-Singer, which is a unified framework designed to have the capability in both visual acoustic matching and stereo singing voice synthesizing.
%, as shown in Fig.~\ref{fig:pipelines}. 
Specifically, VS-Singer comprises a modal interaction network (MIN), a decoder based on the consistency Schrödinger bridge (CSB) and a spatially-aware feature enhancement module (SFE).
% The role of the modal interaction network is to extract spatial features from visual scenes and fuse them with text encoding to produce a linguistic representation.
% The role of the modal interaction network is to analyze the audio-visual consistency through the interaction of different modal information and extract spatial features to guide stereo synthesis.
% Moreover, to address the challenge of slow sampling speed, this paper proposes the CSB, which establishes a tractable diffusion bridge \cite{shi2024diffusion,chen2021likelihood,tang2024simplified} between the representation and the left and right channel mel-spectrograms, enabling the model to generate high-quality stereo singing voices in just one step.
The role of the modal interaction network is to analyze audio-visual consistency through the interaction of different modal information and extract spatial features to guide stereo synthesis.
Moreover, the CSB is also creatively proposed to generate high-quality stereo singing in one step without the need for a teacher model, which increases the synthesis speed to 2 times that of the baseline model.
Additionally, the SFE is employed to enhance the consistency of audio-visual matching. 
Our approach is validated through extensive experiments on the open-source Opencpop \cite{wang2022opencpop} and NVAS-SoundSpace \cite{chen2023novel} corpora. 
Results demonstrate VS-Singer’s superior capability in inference speed and immersive stereo singing voice synthesis.
Audio samples are available at: \url{https://usinger1.github.io/VS-Signer/}.

The key contributions of our work are summarized as follows: 
\begin{itemize}
\item To the best of our knowledge, VS-Singer is the first to integrate visual-acoustic matching and singing voice synthesis into a unified framework for synthesizing stereo singing voice
with room reverberation.
\item We propose a method that fuses Schrödinger bridge with consistency training \cite{song2023consistency} that reduces the cost of training while boosting performance of the model and the speed of inference.

\begin{figure*}[t]
\begin{minipage}[b]{1\linewidth}
  \centering
  \centerline{\includegraphics[width=16cm]{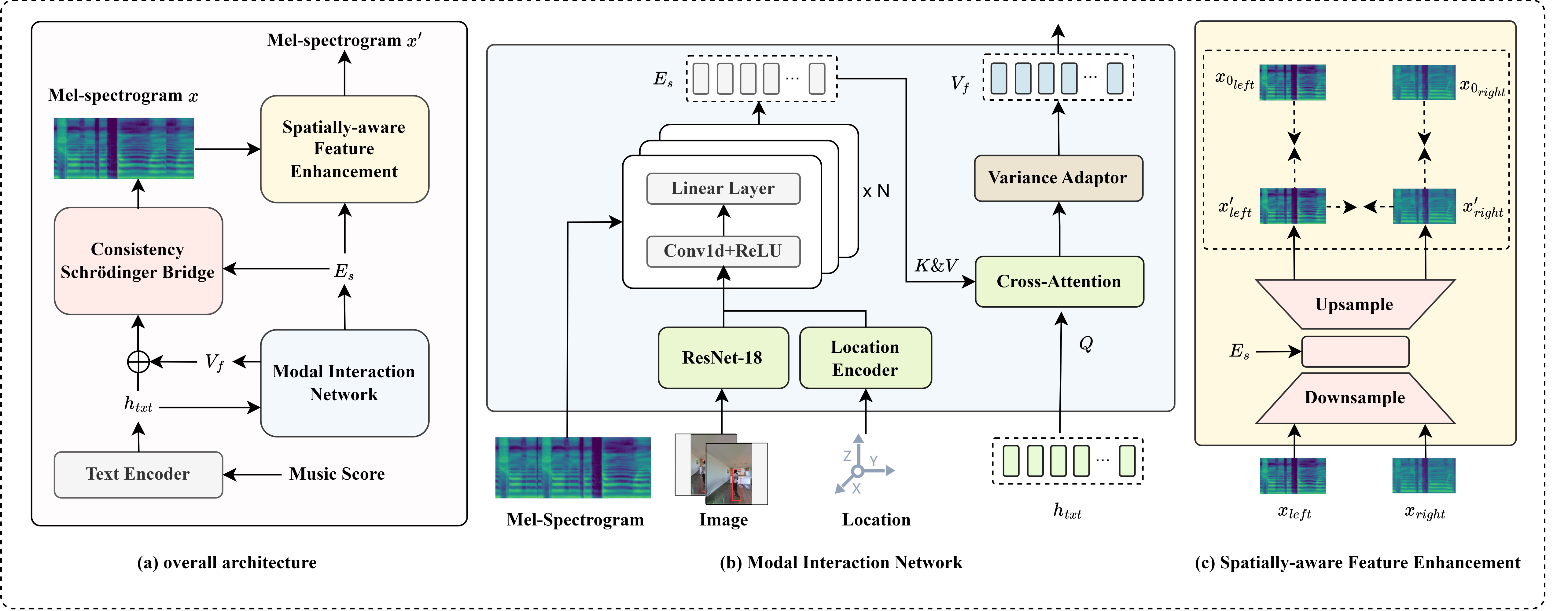}}
%  \vspace{2.0cm}
\end{minipage}
% \vspace{-0.2cm} %%缩减间距
\caption{\fontsize{9pt}{9pt}\selectfont Overview of our proposed model structure. }
\label{fig:overview}
% \vspace{-0.5cm}
\end{figure*}

\item Extensive experiments conducted on open-source corpora demonstrate that VS-Singer can efficiently generate stereo singing voices that accurately matches the scene perspective in one step, particularly achieving a 2x inference speed-up over the comparative cascaded systems.
\end{itemize}

\section{Methodology}
The overall architecture of the proposed VS-Singer model is illustrated in Fig.~\ref{fig:overview}. 
Our model consists of three parts:
%1) The MIN enables the interactive aggregation of multimodal information sources, models environmental acoustics, generates spatial cues, interacts with text encoding, and generates text hidden sequences with spatial features to guide stereo singing generation. 2) The decoder based on consistent Schrödinger bridge establishes optimal transmission and generates stereo samples in a single step. 3) The SFE module enhances the audio difference between left and right channels and accelerates model convergence.
1) The MIN is used to enable the interactive aggregation of multimodal information sources, models environmental acoustics, generates spatial cues, interacts with text encoding, and generates text hidden sequences with spatial features to guide stereo singing generation.
2) The decoder based on CSB is used to establish optimal transmission and generates stereo samples in a single step.
3) The SFE module plays the role of enhancing the audio difference between left and right channels and accelerates model convergence.

\subsection{Modal Interaction Network}
For visual acoustic matching, it is important to consider the different contributions of different regions of the space to the acoustics. Therefore this paper proposes mimicking the observer's natural focus on the scene to infer the effect on acoustics.
Firstly, mask off the 1/4 region of the left and right sides of the scene image to simulate the left and right eye viewpoints, respectively, and extract the corresponding viewpoint spatial feature  $V_{\rm{env}}=(V_{\rm{left}},V_{\rm{right}})$ using the pre-trained ResNet-18 \cite{he2016deep}, where \( (\cdot, \cdot) \) denotes the concatenate operation. Then add the absolute position coding to the visual coding. 
% where \( (\cdot, \cdot) \) denotes the concatenate operation.
Because a single 2D image is insufficient for the model to understand the propagation paths of sound waves in space. Therefore, 3D spatial positioning information is introduced to capture the relative position between the target speaker and source view.
%Therefore, we incorporated 3D spatial positioning information. 
The target speaker's position is recorded as translations along the \(x\), \(y\), and \(z\)  axes. The format is represented as :
% $V_{\rm{loc}} = (d, \sin(\alpha), \cos(\alpha))$
\begin{equation}
    V_{\rm{loc}} = (d, \sin(\alpha), \cos(\alpha)),
\end{equation}
, where \(d\) represents the Euclidean distance between the target speaker and the source viewpoint, and \(\alpha\) is the rotation angle between the target speaker and the source view on the \(XY\) plane. 

The energy of each short-time Fourier transform frame is calculated by calculating the L2-norm of the amplitude of each frame of the binaural channel and quantizing it on a logarithmic scale, which is encoded to generate an energy vector \(V_e\).
The \(V_{env}\), \(V_{loc}\) and \(V_e\) are fed into the modal interaction network to obtain spatial information embedding \(E_s\) aligned with the text. The network includes a convolution stack and a cross-modal attention function.
% The convolution stack fuses the energy embedding with the spatial information at the frame level.
The convolution stack includes convolution layers and pooling layers, which can fuse energy embeddings with spatial information at the frame level and learn the energy changes of stereo at different viewpoints.
The cross-modal attention function enables the model to pay attention to different image region features and spatial information, and infer how they affect reverberation and stereo.

In order to learn the interaction between text encoding and spatial features, an attention mechanism is introduced to compute the dot product of text-hidden sequences and spatial information embedding \(E_s\):
\begin{equation}
    \rm{Attention}(Q, K, V) = \text{softmax}\left(\frac{Q K^T}{\sqrt{d_k}}\right) V,
\end{equation}
where \(Q\) is the text-hidden sequences \(h_{txt}\), and \(K\) and \(V\) is the spatial information embedding.
% This cross-modal attention allows the model to attend to different image region features and spatial information and reason about how they affect the reverberation.
Finally the attention score obtained by the variance adapter\cite{ren2020fastspeech} is projected to the text-hidden sequences to obtain spatially relevant text-hidden sequences \(V_f\).

\subsection{Consistency Schrödinger Bridge}

To ensure the speed and quality of singing voice, we introduce the CSB in this subsection. The consistency model supports one-step sampling and can trade computation time for improved quality. 
We adopt consistency training as in \cite{song2023consistency}, which supports fast sampling while reducing the cost of training a teacher model. 
% However, the consequence of consistency training is performance degradation. 
To compensate for the performance loss caused by independent training, we integrate the Schrödinger Bridge (SB)\cite{shi2024diffusion,chen2021likelihood,tang2024simplified} into score-based generative models (SGM) framework, establishing a tractable process between between the linguistic representation \(x_1\) (e.g., the hidden sequence)
and the clean audio of the left and right channels $x_0=(x_{0_{\rm{left}}},x_{0_{\rm{right}}})$, enhancing the accuracy of marginal density distribution calculations in consistency training, and thereby improving the quality of the samples.
% and the clean audio of the left and right channels $x_0=(x_0_{\rm{left}},x_0_{\rm{right}})$, enhancing the accuracy of marginal density distribution calculations in consistency training, and thereby improving the quality of the samples.
According to \cite{chen2021likelihood}, Schrödinger bridge can be established compatible with the SGM framework. 
By using the diffusion model setup, 
this tractable diffusion bridge can be established between a single data point \( x_0 \) and the linguistic representation \( p_B(x_1 | x_0) \): 
\begin{IEEEeqnarray}{c+x*}\label{sb}
    q(x_t|x_0, x_1) = \mathcal{N}(x_t; \mu_t(x_0, x_1), \Sigma(t)^2), 
    \\
    \mu_t = 
    \frac{\overline{\sigma}_t^2}{\overline{\sigma}_t^2 + \sigma_t^2} x_0 + \frac{\sigma_t^2}{\overline{\sigma}_t^2 + \sigma_t^2} x_1,\quad
    \Sigma(t)^2 = \frac{\overline{\sigma}_t^2 \sigma_t^2}{\overline{\sigma}_t^2 + \sigma_t^2} \cdot I \nonumber,
\end{IEEEeqnarray}
where \(\sigma_t^2 := \int_0^t \beta_\tau d\tau\) and \(\overline{\sigma}_t^2 := \int_t^1 \beta_\tau d\tau\) are variances accumulated from either sides and \(I\) represents the identity matrix. 

When \( x_0 \) derived by SGM from the sampled \( x_t \) of \( p_B \) approaches the clean audio \( x_0 \) asymptotically, the marginal density of the Schrödinger bridge path will coincide with the marginal density induced by diffusion model\cite{liu20232}. 
\cite{song2021scorebased} proposed that any SDE corresponds to a probabilistic flow ordinary differential equation (PF-ODE), whose sampling trajectory at t has the same marginal probability distribution \( p_t(x_t) \).
This means that, given score function \( \nabla \log p_t(x_t) \), the state at any time on the ODE trajectory can be obtained through an ODE solver.
The unbiased estimator of \( \nabla \log p_t(x_t) \) is desigened as:
\begin{equation}\label{cmpt}
\nabla \log p_t(x_t) = - \mathbb{E}\Big[\frac{x_1 - x_t}{\Sigma(t)^2}|x_t \Big],
\end{equation}
% From \eqref{sbsde} and \eqref{cmpt}, 
where \( x_t \) is obtained from \eqref{sb}. 
Correspondingly, the PF-ODE of CSB can be designed as:
\begin{equation}
    dx_t=\Big[f(x_t, t) + \frac{1}{2} \beta(t)^2 \frac{x_1-x_t}{\Sigma(t)^2} \Big]dt.
\end{equation}

To estimate the score, this paper trained a model \( f_\theta(x_t, t, V_f) \), which can map any state on the same PF-ODE trajectory to the same initial state \( x_0 \). 

Specifically, given a data point \( x \), a pair of adjacent data points (\( x_{t_n}, x_{t_{n+1}} \)) can be generated on the same PF-ODE trajectory. Subsequently, the parameters \(\theta\) will be updated through the difference between the Exponential Moving Average (EMA) model\cite{song2023consistency}.
% and the current model's input:
The loss of CSB can be defined as: 
\begin{equation}
\begin{aligned}
& L^N_{CSB}(\theta, \theta^-):= 
\\
& \quad \quad \lambda(t_n) d(f_\theta(x_{t_{n+1}}, t_{n+1}, V_f,E_s), f_{\theta^-}(x_{t_n}, t_n, V_f,E_s)),
\end{aligned}
\end{equation}
where the distance \( d(\cdot, \cdot) \) is calculated using the LPIPS loss  \cite{zhang2018unreasonable}, \( \lambda(\cdot) \in \mathbb{R}^+ \) is a positive weighting function, \(f_{\theta ^-}\) is the target network, and \(f_{\theta}\) is the online network.

In the generation process, the Gaussian noise \(x_n\) and the \(V_f\), \(E_s\) obtained by the spatial interaction fusion module are directly input into the decoder, and the initial data point \(x_0\) is directly calculated from \(x_t\) at any time on the ODE trajectory.

\subsection{Spatially-aware Feature Enhancement}

In order to further enhance the consistency of audio-visual matching, the U-Net network is used to combine the dual-view spatial cues extracted from the modal interaction network with the left and right channel information generated by the CSB. 
This paper proposes to use enhanced loss \(\mathcal{L}_{\rm{enh}}\), that is, to calculate the L2 loss between the the ground truth \(x_0\) and the generated sample audio \(x'\), 
% and the L2 loss between the generated samples, 
reduce the distance between the distance between  enhanced mel and the ground truth that belong to the same channel and expand the distance between different channels.
% to reduce the distance between the audio embedding and the video embedding in the same channel and expand the distance between different channels.
% The output dual-channel mel spectra $x'_{\rm{left}}$ and $x'_{\rm{right}}$ are used to calculate the cross loss $L_{\rm{cro}}$ with the generated sample audio $x_{0_{\rm{left}}}$ and $x_{0_{\rm{right}}}$ to reduce the distance between the distance between  enhanced mel and the ground truth that belong to the same channel:
% and expand the distance between different channels.
\begin{equation}
\mathcal{L}_{\rm{enh}} = || x'_{\rm{left}} - x_{0_{\rm{left}}} ||^2 + || x'_{\rm{right}} - x_{0_{\rm{right}}} ||^2 - || x'_{\rm{left}} - x'_{\rm{right}} ||^2 ,
\end{equation} 

% Finally, the  total mel loss can be expressed as the sum of the enhancement mel loss $L_{\rm{enh}}$ and the previous CSB loss $L_{\rm{CSB}}$:
% \begin{equation}
%     \mathcal{L}_{\rm{mel}}=\mathcal{L}_{\rm{enh}}+
%     \mathcal{L}_{\rm{CSB}}.
% \end{equation}

% This process further enhances binaural audio-visual consistency and accelerating the model's ability to learn the interaction between spatial structures and sound propagation during training.

% To further enhance the consistency of audio-visual matching, 
% the left and right channel mel spectrograms $x_{\rm{left}}$ and $x_{\rm{right}}$ output by the decoder are concatenated and then fed into a U-Net network \cite{zhou2020sep}. Simultaneously, the visual spatial features are expanded to match the dimensionality of the mel spectrogram features, and then the audio and visual spatial feature maps are concatenated along the channel dimension.
% Then we extract the audio embeddings (or energy embeddings) of the left and right channels of the real audio and calculate the cross loss with the generated sample audio to reduce the distance between the audio embedding and the video embedding in the same channel and expand the distance between different channels.

\begin{table*}[t!]
    \centering
    \caption{Results of comparison with the cascaded systems.}
    \begin{tabular}{l|c|ccccc|ccccc}
        \toprule % 顶部横线
        \multirow{2}{*}{\textbf{System}} &
        \multirow{2}{*}{\textbf{NFE}} &
        \multicolumn{5}{c|}{\textbf{Test-Unseen}} &
        \multicolumn{5}{c}{\textbf{Test-Seen}} \\
        &  & MOS $\uparrow$ & RTF $\downarrow$ & MCD $\downarrow$ & LRE $\downarrow$ & RTE $\downarrow$ & MOS $\uparrow$ & RTF $\downarrow$ & MCD $\downarrow$ & LRE $\downarrow$ & RTE $\downarrow$  \\
        \midrule % 中间的分隔横线
        GT & - & 4.35 $\pm$ 0.12 & - & -& - & - & 4.32 $\pm$ 0.18 & - & - & - & - \\
        GT(voc.) & - & 4.23 $\pm$ 0.16 & - & 1.85 & 0.154& 0.005 & 4.19 $\pm$ 0.15 & - & 1.82 & 0.153 & 0.005 \\
        \midrule % 中间的分隔横线
        DiffSinger & 50 & 3.62 $\pm$ 0.11 & 0.203 & 7.89 & 0.945 & 0.073 & 3.68 $\pm$ 0.09 & 0.208 & 7.88 & 0.936 & 0.068 \\
        VISinger2 & - & 3.72 $\pm$ 0.06 & 0.069 & 7.81 & 0.939 & 0.069 & 3.79 $\pm$ 0.10 & 0.069 & 7.74 & 0.932 & 0.064 \\
        CoMoSpeech & 1 & 3.46 $\pm$ 0.08 & 0.052 & 8.61 & 0.951 & 0.070 & 3.51 $\pm$ 0.07 & 0.055 & 8.53 & 0.942 & 0.063 \\
        CoMospeech & 4 & 3.70 $\pm$ 0.09 & 0.063 & 7.74 & 0.948 & 0.068 & 3.76 $\pm$ 0.12 & 0.064 & 7.72 & 0.933 & 0.062 \\
        Ours & 1 & 3.48 $\pm$ 0.12 & \textbf{0.024} & 8.55 & 0.982 & 0.077 & 3.56 $\pm$ 0.10 & \textbf{0.021} & 8.49 & 0.968 & 0.069 \\
        Ours & 4 & \textbf{3.74} $\pm$ \textbf{0.13} & 0.033 & \textbf{7.71} & \textbf{0.907} & \textbf{0.065} & \textbf{3.81} $\pm$ \textbf{0.08} & 0.030 & \textbf{7.65} & \textbf{0.889} & \textbf{0.058} \\
        \bottomrule % 底部的横线
    \end{tabular}
    % \caption{Caption}
    \label{table:comparison}
\end{table*}

\begin{table}
    \centering 
    % \normalsize
    \scriptsize
    \caption{Ablation Study on the Opencpop and NVAS-SoundSpace datasets.}
    \resizebox{\linewidth}{!}{
    \begin{tabular}{lcccc}
      \toprule
      \textbf{Model} & \textbf{ MOS} $\uparrow$ & 
          \textbf{MCD} $\downarrow$ &
      \textbf{LRE} $\downarrow$    
      & \textbf{RTE} $\downarrow$   \\
      % \cline{1-5}
      \midrule
      VS-Singer & \textbf{3.81} $\pm$ \textbf{0.08}
      %\textbf{3.81} $\pm $ \textbf{0.14} 
      & \textbf{7.65}  & \textbf{0.889} & \textbf{0.058}
      \\
       % \cline{1-5}
       \midrule
      \quad w/o MIN & 3.65 $\pm$ 0.11 & 7.82  & 1.257 & 0.084\\
      \quad w/o SB   & 3.50 $\pm$ 0.06 & 8.63  & 0.899 & 0.062 \\
      \quad w/o SFE & 3.72 $\pm$ 0.10 & 7.72  & 0.921 & 0.069 \\
      
      % \hline
      \bottomrule
    \end{tabular}}
    \label{table:ablation}
    % \vspace{-0.45cm}
\end{table}

\section{Experiments}
\subsection{Data Preparation}

This paper evaluated the proposed method using the public datasets Opencpop\cite{wang2022opencpop} and NVAS-SoundSpace\cite{chen2023novel}. The Opencpop dataset contains 100 Chinese songs performed by a female singer. 
% The NVAS-SoundSpace dataset, generated using the SoundSpaces 2.0 platform, includes multi-view binaural audio-visual data of two people conversing in 3D scenes. This dataset consists of 1,000 speakers, 120 3D scenes, and 200K viewpoints, resulting in 13,000 hours of binaural audio-visual data. 
The NVAS-SoundSpace dataset contains 13,000 hours of binaural audiovisual data.
The average difference was calculated between the left and right channel audio and selected videos where this difference exceeded 0.01 to identify qualifying binaural audio. Subsequently, this paper applied convolutional reverberation by transferring the impulse responses extracted from NVAS-SoundSpace audio to the Opencpop audio, generating binaural audio. 
Following the data split method proposed by DiffSinger\cite{liu2022diffsinger}, 95 songs were selected as the training set and 5 songs as the validation set.
According to the experimental scheme of AViTAR\cite{chen2022visual}, the test set is divided into ``test-seen" and ``test-unseen".

\subsection{Experimental Setup}

In our experiments, all audio data were resampled to 22,050 Hz, with the hop size set to 128 and the frame size set to 512. The number of mel bins \(H_m\) was 80. The mel-spectrograms were linearly scaled to the range \([-1, 1]\), and the \(F_0\) was normalized to have zero mean and unit variance. The deep neural network was trained on a single NVIDIA 4090 GPU using the Adam optimizer, with an initial learning rate of \(10^{-4}\) and a batch size of 16, over a total of 800,000 steps. The maximum number of discrete steps \(N\) was set to 120. We set the minimum processing time \(\epsilon = 0.001\) and the maximum processing time \(T = 0.999\) to ensure numerical stability. ResNet-18\cite{he2016deep}, pretrained on ImageNet\cite{deng2009imagenet}, was used as the visual encoder to extract visual features. 

Since there is currently no model that directly synthesizes stereo audio with room reverberation through visual guidance, this paper chose to evaluate the effectiveness of our proposed method by comparing it with a cascade of models: a singing voice synthesis model, a vision-guided environmental acoustic matching model, and a vision-guided mono-to-stereo conversion model. Specifically, the singing voice generated by the synthesis model is processed through the environmental acoustics matching model to add room reverberation, and finally converted into binaural audio.  
In the experimental results, we will only use the singing synthesis model to represent the cascade model.
The cascaded baseline is constructed based on the following models: 
1) GT, the ground truth singing audio. 2) GT(voc.), convert GT's mel-spectrograms back to audio using vocoder. 3) DiffSinger\cite{liu2022diffsinger}, a singing voice synthesis model based on diffusion model. 4) VISinger2\cite{zhang2022visinger}, an end-to-end singing voice synthesis model based on VITS\cite{kim2021conditional}. 5) CoMoSpeech\cite{ye2023comospeech}, a speech synthesis model using consistency distillation\cite{song2023consistency} with DiffSinger\cite{liu2022diffsinger} as the teacher model. 6) LeMARA\cite{DBLP:conf/nips/SomayazuluCG23}, a visual acoustic matching model that adds room reverberation to a given audio based on images. 7) Sep-Stereo\cite{zhou2020sep}, a model that converts mono audio to binaural audio using scene images.

\subsection{Experimental Results and Analysis}

In the objective evaluation, this paper measures performance from four aspects: 
1) Real-Time Factor (RTF), the time required for the system to synthesize one second waveform.
2) Mel Cepstral Distortion (MCD), this metric measures the degree of Mel-cepstral distortion between the synthesized speech and the reference speech. 
3) Left-Right Energy Ratio Error (LRE), evaluates the correctness of spatial sound by measuring the difference between the energy ratio of the left and right channels. 
4) RT60 Error (RTE)\cite{chen2022visual}, assesses the correctness of acoustic properties by measuring the reverberation time error for a 60dB decay (RT60). 
Number of function evaluation (NFE) is the total number of times the denoiser function is evaluated during the generation process.
For the subjective evaluation, to evaluate the naturalness of the synthesized speech, 30 native Chinese speakers were asked to rate the samples using MOS on a scale of 1 to 5.

From Table~\ref{table:comparison}, it can be seen that although the values of each metric in the unseen scene are slightly lower than those in the seen scene, our model achieves the best results in each metric.
Compared with the baseline model, our model has significantly lower RTF. 
This mainly attributed to the fact that the consistency model can generate in just one step.
Moreover, our model achieved higher MOS and lower MCD than CoMoSpeech, which uses consistency distillation as singing synthesis part, demonstrating that the introduction of the Schrödinger bridge effectively addresses the performance degradation associated with independently training a consistency model. 
Additionally, it also can be observed that our model possesses excellent spatial awareness, as evidenced by the lowest LER and RTE values. 
All results indicate that our model can effectively synthesize speech with corresponding room reverberation based on the spatial information provided by the input images, without compromising speech quality.

\subsection{Ablation Study}

\begin{figure}[t]

\begin{minipage}[b]{1\linewidth}
  \centering
  \centerline{\includegraphics[width=1\textwidth,height=0.72\textwidth]{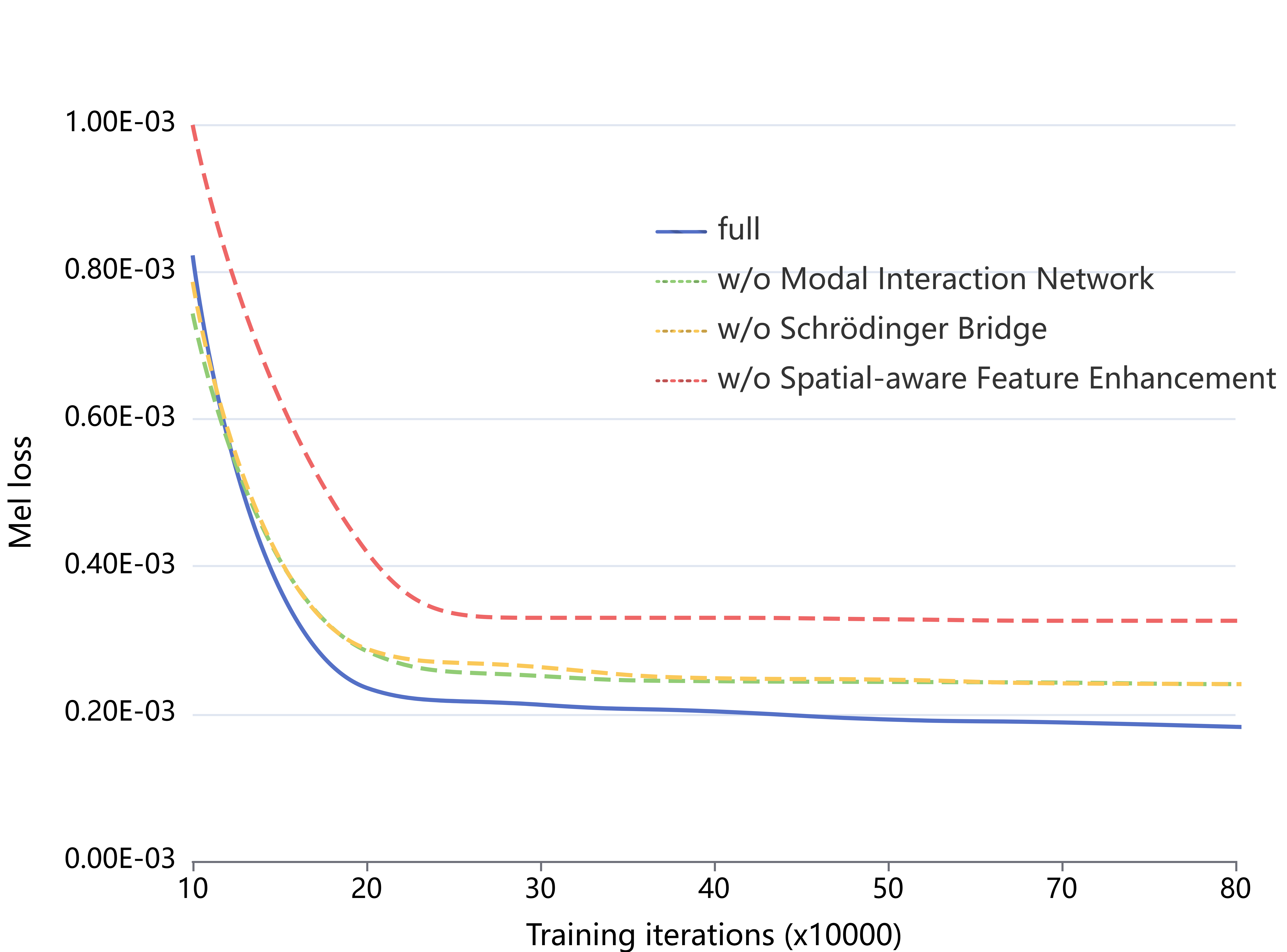}}
%  \vspace{2.0cm}
\end{minipage}

% \vspace{-0.2cm} %%缩减间距
\caption{\fontsize{9pt}{9pt}\selectfont Convergence speed of the models at 800k training steps in the ablation experiment.}
\label{fig:abl}
% \vspace{-0.5cm}
\end{figure}

In this subsection, an ablation analysis is conducted to evaluate the impact of each module on the model's performance. 
From Table~\ref{table:ablation}, it can be observed that removing MIN leads to a sharp increase in RTE and LRE, indicating that MIN effectively helps the model understand the positioning of characters in the scene. 
The removal of the SFE module shows a similar trend. Fig.~\ref{fig:abl} demonstrates that this module accelerates model convergence and
aids the model in further perceiving the spatial structure of the scene. 
As expected, the removal of the Schrödinger bridge while retaining the consistency model results in the lowest MOS and the highest MCD, which suggests that the Schrödinger bridge compensates for the performance degradation caused by training the consistency model independently.
All these results indicate the effectiveness of each component in the model.

\section{Conclusion}

This paper proposes VS-Singer, a novel framework that unifies spatial acoustic matching and stereo singing voice generation.
VS-Singer consists of a modal interaction network, a decoder based on consistency Schrödinger bridge and a spatially-aware feature enhancement module.
The modal interaction network is introduced to add the spatial information into hidden text sequences.
% Leveraging the text hidden sequence as linguistic representation, 
Then, a tractable consistency Schrödinger bridge between the linguistic representation and the mel-spectrogram of the binaural channels is established, which boosts performance and speed while reducing training costs.
Furthermore, the spatially-aware feature enhancement module can improve its spatial perception ability.
Extensive experiments on open-source corpora demonstrate that the proposed method outperforms cascaded systems in both stereo singing voice quality and synthesis speed.

\bibliographystyle{IEEEtran}
\bibliography{mybib}

\end{document}